\documentclass[prb,twocolumn,amssymb,showpacs,superscriptaddress]{revtex4}
\usepackage{amsmath} 
\usepackage{amssymb} 
\usepackage{graphicx} 
\usepackage{epsfig}
\usepackage{epstopdf}
\usepackage{color}
 \usepackage{rotating}

\begin{document}
 \title{Self-energy enhancements in doped Mott insulators}
\author{J. Merino}
\affiliation{Departamento de F\'isica Te\'orica de la Materia Condensada, Condensed Matter Physics Center (IFIMAC) and
Instituto Nicol\'as Cabrera, Universidad Aut\'onoma de Madrid, Madrid 28049, Spain}
\author{O. Gunnarsson}
\affiliation{Max-Planck Institute f\"ur Festk\"orperforschung, Heisenbergstra$\beta$e 1, D-70569 Stuttgart, Germany}
\author{G. Kotliar}
\affiliation{Department of Physics and Astronomy, Rutgers University, Piscataway, New Jersey 08854, USA}
\date{\today}
\begin{abstract}
We analyze enhancements in the magnitude of the self-energy for electrons far away from the Fermi surface in doped Mott insulators using the 
dynamical cluster approximation to the Hubbard model. For large onsite repulsion, $U$, and hole doping,  the magnitude of the self-energy 
for imaginary frequencies at the top of the band (${\bf k}=(\pi,\pi)$) is enhanced with respect to the self-energy magnitude at the bottom of the 
band (${\bf k}=(0,0)$). The self-energy behavior at these two ${\bf k}$-points is switched for 
electron doping. Although the hybridization is much larger for $(0,0)$ than for $(\pi,\pi)$, we demonstrate that this is not the
origin of this difference.  Isolated clusters under a downward shift of the chemical potential, $\mu<U/2$, at half-filling reproduce the 
overall self-energy behavior at $(0,0)$ and $(\pi,\pi)$ found in low hole doped embedded clusters. This happens although there is 
no change in the electronic structure of the isolated clusters. Our analysis shows that a  downward shift of the chemical potential
which weakly hole dopes the Mott insulator can lead to a large enhancement of the $(\pi,\pi)$ self-energy  which is not necessarily associated 
with electronic correlation effects, even in embedded clusters. 
\end{abstract}
\pacs{71.30.+h; 71.27.+a; 71.10.Fd}
\maketitle 

\section{Introduction}

Understanding the electronic properties of two-dimensional metals close to the Mott insulator \cite{mott,imada} 
transition remains a formidable theoretical challenge. A remarkable example is found in the cuprates in which 
high-Tc superconductivity arises when doping the Mott insulator\cite{lee}.  Although these systems have been 
studied intensively over the past decades there is a lack of consensus on the mechanism of superconductivity. The 
simplest electronic correlated model which can capture the electronic properties and phase diagram of the cuprates 
is the Hubbard model on a square lattice. Recent progress in numerical approaches \cite{jarrell2000,kotliar2001,jarrell2005} to
strongly correlated electron systems allows for an accurate determination of the electron spectra of the Hubbard model even in the relevant but difficult regime 
of a large onsite Coulomb repulsion, $U$. The electron spectra obtained from these approaches can be compared with 
ARPES experiments \cite{ARPES} testing the validity of the model.

The self-energy in imaginary frequencies, $\Sigma_{\bf k}(i\omega_n)$, is the key quantity encoding the strength of electron 
correlations. determining the Greens function through Dyson's equation:
\begin{equation}
G_{\bf k}(i \omega_n)={1 \over i\omega_n +\mu -\epsilon_{\bf k}-\Sigma_{\bf k}(i\omega_n)},
\label{eq:green}
\end{equation}
where: $\omega_n={(2 n+1) \pi  \over \beta}$, are Matsubara frequencies, $n$ is an integer,  and $\mu$ 
the chemical potential,  $\beta$ the inverse of the temperature, $\beta=1/T$. Deviations from independent electron 
behavior due to Coulomb interactions can be monitored through the quasiparticle weight, $Z_{\bf k}$, obtained from\cite{tremblay}:
\begin{equation}
 Z_{\bf k}={1 \over 1-{\text{Im} \Sigma_{\bf k}( i\omega_n) \over \omega_n}  } \bigg|_{\omega_n \rightarrow 0}.
\end{equation}
In the dynamical mean field theory (DMFT), the magnitude of the local self-energy, $|\text{Im}\Sigma(i\omega_n)|$, is enhanced 
with $U$ until\cite{kotliar1996,kotliar2006} Im$\Sigma(i\omega_n)|_{\omega_n\rightarrow 0} \rightarrow -\infty$ as $U \rightarrow U_{c2}$, 
the critical value for the Mott transition from the metallic phase. Quasiparticles disappear\cite{merino2008} uniformly, $Z_{\bf k}=Z \rightarrow 0$, and 
a Mott-Hubbard gap opens over the whole Fermi surface. This scenario is modified by non-local electron correlations in cluster extensions of DMFT such as 
cellular-DMFT (CDMFT)\cite{kotliar2001,park2008,liebsch2009} and the dynamical cluster approximation\cite{jarrell2005} (DCA) which rather 
obtain anisotropic self-energy enhancements over the Fermi surface with larger  $|\text{Im}\Sigma_{\bf K}(i\omega_n)|$ in the antinodal region around
the coarse-grained momentum: ${\bf K}=(\pi,0)$, than the nodal region for ${\bf K}=(\pi/2,\pi/2)$,\cite{civelli2005,gull2009,jarrell2005,macridin,gull2010,sordi2012,merino2014}. This leads to 
a pseudogap \cite{jarrell2005,gull2013,sordi2012} in the electron spectra consistent with ARPES experiments on cuprates \cite{ARPES}. The origin of the pseudogap at $(\pi,0)$
in embedded cluster calculations has unambiguously been identified with spin fluctuations  \cite{kyung,macridin}  based on the 
fluctuation diagnostics approach \cite{gunnarsson2015}. 

Electronic correlation effects at the Fermi surface are then signaled by 
large enhancements of  $|\text{Im}\Sigma_{\bf K}(i\omega_n)|$  regardless of their origin.  
In order to fully characterize the ground state and excitations of the Hubbard model it is useful
to quantify the strength of electron correlations not only at the Fermi surface but also in regions of the first
Brillouin zone which are away from it.  Recent work has reported large self-energy enhancements 
far away from the Fermi surface\cite{maier2002, civelli2005,kotliar2007,gull2010} in hole doped Mott insulators.   At low 
hole dopings and large-$U$,  $|\text{Im}\Sigma_{\bf (\pi,\pi)}(i\omega_n)|$ can be comparable or even {\it larger} than 
$|\text{Im}\Sigma_{(\pi,0)}(i\omega_n)|$ and much larger than $|\text{Im}\Sigma_{(0,0)}(i\omega_n )|$. 
Motivated by these recent findings we focus on self-energies far away from the Fermi surface.
More specifically we would like to understand the origin of the large enhancement in
$|\text{Im}\Sigma_{\bf (\pi,\pi)}(i\omega_n)|$ found in embedded clusters:  do self-energy enhancements away from the
surface necessarily correspond to electronic correlation effects?.  Our study adds relevant information to previous works 
which have concentrated on electronic correlation effects at the Fermi surface. We first corroborate that the magnitude of the self-energy
at $(\pi,\pi)$ is larger than at $(0,0)$ in low hole doped Mott insulators  in agreement with 
previous works\cite{maier2002,kotliar2007}. We have analyzed the origin of such behavior finding how the large differences 
in bath hybridization functions for different {\bf k}-vectors are not responsible for the difference between $(\pi,\pi)$ and $(0,0)$ self-energies. We then analyze 
self-energies of isolated clusters finding that the self-energy behavior  observed in DCA calculations also occurs in isolated clusters. A downward 
shift of the chemical potential in an isolated cluster with fixed occupation, $n=1$, leads to larger $|\text{Im}\Sigma_{(\pi,\pi)}( i\omega_n)|$ 
than $|\text{Im}\Sigma_{(0,0)}( i\omega_n)|$ as observed in DCA calculations of embedded clusters. From the frequency dependence of  
$(0,0)$ and $(\pi,\pi)$ spectral densities, we conclude that the different self-energy behavior does not correspond 
to stronger correlation effects acting at $(\pi,\pi)$ than at $(0,0)$ but is a consequence of breaking particle-hole 
symmetry of the isolated cluster.  In electron doped Mott insulators the situation is switched:  
$|\text{Im}\Sigma_{(0,0)}( i\omega_n)|>|\text{Im}\Sigma_{(\pi,\pi)}( i\omega_n)|$ since $\mu > U/2$ in this case.

We briefly introduce the model in Sec. \ref{sec:model}. In Sec. \ref{sec:DCA} we analyze how hole doping the Mott insulator off half-filling 
modifes the DCA self-energies. We also demonstrate the negligible role played by bath-cluster hybridizations on the different behavior of  $(0,0)$ and $(\pi,\pi)$ self-energies. 
In Sec. \ref{sec:isol} we show how such difference is already present in isolated clusters and can be attributed to 
breaking particle-hole symmetry. The implications of our analysis to self-energy enhancements observed far away from the Fermi surface in
embedded clusters are finally discussed in Sec. \ref{sec:conclusion}. 

\section{Model and formalism}
\label{sec:model} 
 
We consider the Hubbard model on a square lattice:
\begin{equation}
H=t\sum_{\langle ij \rangle} (c^\dagger_{i\sigma} c_{j\sigma} +c^\dagger_{j\sigma} c_{i\sigma})+ U\sum_i n_{i\uparrow}n_{j\downarrow}
-\mu\sum_{i\sigma}n_{i\sigma},
\end{equation}
where $t$ is the nearest neighbors hopping integral and $U$ the onsite Coulomb repulsion and 
$n_{i\sigma}=c^\dagger_{i\sigma}c_{i\sigma}$. We analyze the half-filled, $n=1$, and the 
doped Hubbard model relevant to the cuprates.

The model is solved using DCA from which the different self-energies\cite{jarrell2005} are 
obtained. We consider $N_c=4,8$ clusters embedded in a self-consistent bath. The quantum cluster problem 
is solved using the Hirsch-Fye algorithm\cite{hirsch}.

 \section{Self-energy enhancements in DCA}
\label{sec:DCA}
 
\begin{figure}
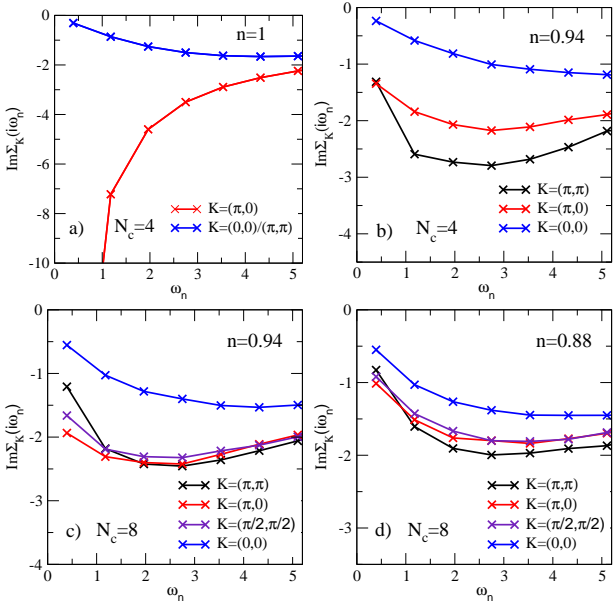

\epsfig{file=fig1a.eps,width=4cm,clip=}
\epsfig{file=fig1b.eps,width=4cm,clip=}
\epsfig{file=fig1c.eps,width=4cm,clip=}
\epsfig{file=fig1d.eps,width=4cm,clip=}
\caption{\label{fig:fig1} (Color online) DCA self-energies of the Hubbard model on the square lattice. DCA results for $N_c=4$ on the half-filled model, $n=1$, are shown in a)
which are compared with the weakly doped case, $n=0.94$ in b). Results for hole doped $N_c=8$ clusters are shown in  c) for $n=0.94$ and in d)  for $n=0.88$. 
The parameters used in the calculations are: $t=-1$, $U=8$ and $\beta=8$.}
\end{figure}

 We have performed DCA calculations on $N_c=4,8$ clusters for $t=-1$, $U=8$ and $\beta=8$. 
The DCA self-energies obtained for $N_c=4$ are shown in Fig. \ref{fig:fig1}. For $n=1$, the  ${\bf K}=(\pi,0)/(0,\pi)$ 
self-energies  display divergent behavior, Im$\Sigma_{(\pi,0)}(i\omega_n)|_{\omega_n \rightarrow 0}  \rightarrow -\infty $, due to  the opening of a 
Mott-Hubbard gap. The $(0,0)$ and $(\pi,\pi)$ cluster self-energies are identical due to particle-hole symmetry 
and at low frequencies: Im$\Sigma_{(0,0)/(\pi,\pi)}(i\omega_n)|_{\omega_n \rightarrow 0}  \rightarrow 0$. Under weak hole doping, $n=0.94$, such divergence 
disappears and Im$\Sigma_{(\pi,0)/(0,\pi)}(i\omega_n)|_{\omega_n \rightarrow 0} \rightarrow 0$ as
$\omega_n \rightarrow 0$.  On the other hand, $(0,0)$ and $(\pi,\pi)$ become inequivalent so that
$|\text{Im}\Sigma_{(\pi,\pi)}(i\omega_n)|>|\text{Im}\Sigma_{(0,0)}(i\omega_n)|$. Such asymmetry is robust against an 
increase in the size of the cluster as shown in Fig. \ref{fig:fig1} c) and d) for $N_c=8$. 

\begin{figure}
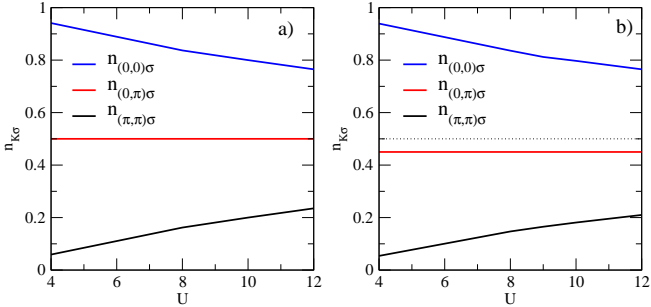

\epsfig{file=fig2a.eps,width=4.25cm,clip=}
\epsfig{file=fig2b.eps,width=4.25cm,clip=}
\caption{\label{fig:fig2} (Color online) Dependence of sector occupations on $U$ from DCA calculations for $N_c=4$ clusters. 
We show results of $n_{{\bf K}\sigma}$ for the undoped model, $n=1$ in a) and for low dopings, $n=0.94$, in b).  
The horizontal dashed line denotes a half-filled sector, $n_{{\bf K}\sigma}=1/2$. 
We have used: $t=-1$, $U=8$, and $\beta=8$.  }
\end{figure}

In order to elucidate the origin of the differences between the $(0,0)$ and $(\pi,\pi)$ self-energies
we first analyze the behavior of sector populations. In Fig. \ref{fig:fig2} the sector populations per spin, $n_{{\bf K}\sigma}$, in a half-filled system, $n=1$, are compared with 
the low hole doped system: $n=0.94$. As expected, the doped holes mainly populate the $(\pi,0)/(0,\pi)$ sectors:  $n_{(\pi,0)\sigma}=n_{(0,\pi)\sigma}<0.5$ since these are closest to the Fermi energy. 
On the other hand the $n_{(\pi,\pi)\sigma}$ and $n_{(0,0)\sigma}$ populations are weakly affected by doping as shown in Fig. \ref{fig:fig2}. Since these populations satisfy:
$n_{(\pi,\pi)\sigma} \approx 1-n_{(0,0)\sigma}$,  we would expect that ${\bf K}=(0,0)$ and ${\bf K}=(\pi,\pi)$ behave in a similar way under hole doping as
for $n=1$ shown in Fig. \ref{fig:fig1} a).
Hence, from the values of the $n_{(\pi,\pi)}$ and $n_{(0,0)}$ populations under hole doping we would have expected that for $n=0.94$, 
$\text{Im}\Sigma_{(0,0)}(i\omega_n)\approx \text{Im}\Sigma_{(\pi,\pi)}(i\omega_n)$ in contrast to the DCA results plotted in Figs. \ref{fig:fig1} b)-d).  
We note that for electron doping the behavior of the $(0,0)$ and $(\pi,\pi)$ 
self-energies is switched so that (not shown): $|\text{Im}\Sigma_{(0,0)}(i\omega_n)|>|\text{Im}\Sigma_{(\pi,\pi)}(i\omega_n)|$.

\begin{figure}
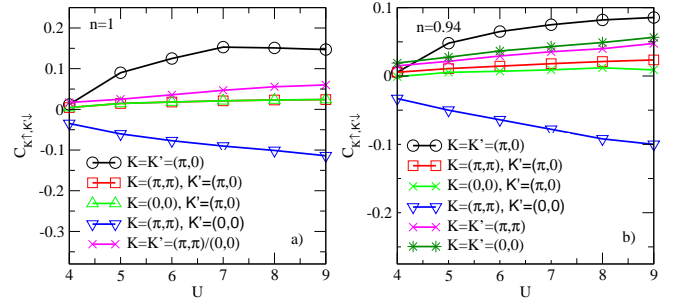

\epsfig{file=fig3a.eps,width=4.25cm,clip=}
\epsfig{file=fig3b.eps,width=4.25cm,clip=}
\caption{\label{fig:fig3} (Color online) Dependence of correlation functions, $C_{{\bf K} \sigma,{\bf K'} \sigma'}$, with Coulomb interaction $U$ as obtained from DCA. 
Results for the half-filled Hubbard model, $n=1$, are shown in a) which can be compared with results for the hole doped,  $n=0.94$, case shown in b). 
We have used: $t=-1$ and $\beta=8$. }
\end{figure}

Changes in ground state properties can be monitored by evaluating correlation functions, $C_{{\bf K} \sigma,{\bf K'} \sigma'}= \langle n_{\bf K \sigma} n_{\bf K' \sigma'} \rangle-
\langle n_{\bf K \sigma}\rangle \langle n_{\bf K' \sigma'} \rangle$. For instance, the opening of a pseudogap in the spectral function  \cite{civelli2005,gull2010,jarrell2005}  
has been related with the formation of an RVB state in the cluster\cite{merino2014} through the dependence of $C_{{\bf K} \sigma,{\bf K'} \sigma'}$ on $U$. The formation of a 
RVB state in the cluster is signaled by $C_{{\bf K} \uparrow,{\bf K'} \downarrow} >0$ with ${\bf K}={\bf K'}=(\pi,0)$ or $(0,\pi)$. 
Here, we analyze the behavior of $C_{{\bf K} \sigma,{\bf K'} \sigma'}$ involving the $(0,0)$ and $(\pi,\pi)$ sectors as shown in Fig. \ref{fig:fig3} for $n=1$ and $n=0.94$.  
For $n=1$, $C_{(0,0) \uparrow, (0,0)\downarrow}=C_{(\pi,\pi) \uparrow, (\pi,\pi)\downarrow}$
and $C_{(\pi,0) \uparrow, (\pi,\pi)\downarrow}=C_{(\pi,0) \uparrow, (0,0)\downarrow}$ as expected from particle-hole symmetry. 
At low hole dopings, this behavior is modified so that $(0,0)$ and $(\pi,\pi)$ become inequivalent since
$C_{(\pi,0) \uparrow, (\pi,\pi)\downarrow} \gtrsim C_{(\pi,0) \uparrow, (0,0)\downarrow}$ and $C_{(0,0) \uparrow, (0,0)\downarrow} \gtrsim C_{(\pi,\pi) \uparrow, (0,0)\downarrow}$.  
However, these differences seem to be too small to explain the corresponding significant differences between the $(0,0)$ and $(\pi,\pi)$ self-energies 
shown in Fig. \ref{fig:fig1}.

The different magnitudes of the $(0,0)$ and $(\pi,\pi)$ self-energies could arise from the different coupling strengths of the two sectors to the bath. 
We now analyze the doping dependence of the bath-cluster hybridization functions, $\Gamma_{\bf K}(i\omega_n)$ for ${\bf K}=(0,0)$ and $(\pi,\pi)$. 
These can be obtained from \cite{merino2014}:
\begin{equation}\label{eq}
 {\rm Im} \Gamma_{\bf K}(i\omega_n)= { {\rm Im} G_{0{\bf K}}(i\omega_n)
\over [{\rm Re} G_{0{\bf K}}(i\omega_n)]^2+[{\rm Im} G_{0{\bf K}}(i\omega_n)]^2}+\omega_n,
\end{equation}
with $G_{0{\bf K}}( i\omega_n)$ the cluster excluded Greens function.
In Table \ref{table1} DCA results for ${\rm Im} \Gamma_{\bf K}(i\omega_n)$ on 
half-filled $N_c=4$ clusters, $n=1$, are compared with hole doped clusters with $n=0.94$. 
At half-filling both $(0,0)$ and $(\pi,\pi)$ sectors are coupled 
with identical strengths  due to particle-hole symmetry. 
\begin{table}
\caption{Bath-cluster hybridization functions, $\Gamma_{\bf K}(i\omega_n)$,  obtained from DCA on $N_c=4$ clusters.
The half-filled $n=1$ and doped $n=0.94$ cases are compared. Parameters are $U=8$, $\beta=8$ and $t=-1$.}
\label{table1}
\begin{tabular}{llll}
n & Im$\Gamma_{(\pi,\pi)}(i\pi/\beta)$ &  Im$\Gamma_{(0,\pi)}(i\pi/\beta)$ & Im$\Gamma_{(0,0)}(i\pi/\beta)$ \\
\hline
1 &  -0.15 & -0.035 & -0.15 \\
0.94 & -0.0423   &  -0.325     &  -0.185   \\
\hline
\end{tabular}
\end{table}
However, for hole doping the $(0,0)$ hybridization is stronger than
the $(\pi,\pi)$ hybridization  as shown in table \ref{table1}.   The asymmetry in
Im$\Gamma_{\bf K}(i\pi/\beta)$ is related to the downward shift of $\mu$ needed for hole doping the system. 
As $\mu$ is decreased below $U/2$, Im$G_{0(\pi,\pi)}(i\pi/\beta)$ is suppressed while Im$G_{0(0,0)}(i\pi/\beta)$ 
is increased from their corresponding values at half-filling. Similar behavior of the hybridization functions is found for $N_c=8$. 
The fact that $|\text{Im}\Gamma_{(0,0)}(i\pi/\beta)|>|\text{Im}\Gamma_{(\pi,\pi)}(i\pi/\beta)|$
could naturally explain the larger magnitude of the $(\pi,\pi)$ self-energy as compared to $(0,0)$ found in DCA. 
 
\begin{figure}
\epsfig{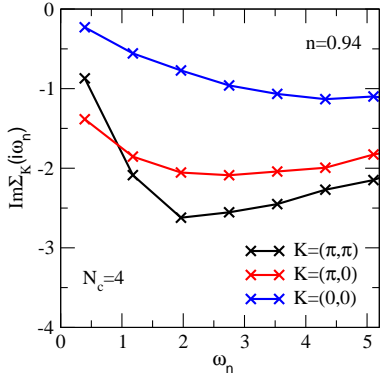}
\caption{\label{fig:fig4} (Color online) DCA self-energies with $(0,0)$ and $(\pi,\pi)$ bath-cluster hybridizations equal on a $N_c=4$ cluster.
The plot shows how the larger $(\pi,\pi)$ self-energy enhancement: $|\text{Im}\Sigma_{(\pi,\pi)}(i\omega_n)|>|\text{Im}\Sigma_{(0,0)}(i\omega_n)|$,  
shown in Fig. \ref{fig:fig1} is present even when we artificially fix: $\Gamma_{(0,0)}(i\omega_n)=\Gamma_{(\pi,\pi)}(i\omega_n)$.
The parameters  are: $t=-1$, $U=8$, $\beta=8$ and $n=0.94$.}
\end{figure}

The role played by the different $(0,0)$ and $(\pi,\pi)$ hybridizations on the corresponding self-energies
can be checked by fixing: $\Gamma_{(0,0)}(i\omega_n)=\Gamma_{(\pi,\pi)}(i\omega_n)$.  The results shown in Fig. \ref{fig:fig4} 
demonstrate how, even in this situation,  $|\text{Im}\Sigma_{(\pi,\pi)}(i\omega_n)|>|\text{Im}\Sigma_{(0,0)}(i\omega_n)|$ as in the actual self-consistent 
DCA calculations containing the different hybridizations of Table \ref{table1}. This result indicates that the larger $(\pi,\pi)$ self-energy enhancement:
$|\text{Im}\Sigma_{(\pi,\pi)}(i\omega_n)|>|\text{Im}\Sigma_{(0,0)}(i\omega_n)|$, observed in DCA calculations 
is not due to the $(\pi,\pi)$ hybridization being weaker than the $(0,0)$ hybridization.  

\section{Self-energy enhancements in isolated clusters}
\label{sec:isol}

We have found above that the coupling to the bath plays a minor role on the different behavior of the $(0,0)$ and
$(\pi,\pi)$ DCA self-energies. Motivated by this fact we have decoupled the cluster from the bath and have analyzed 
the self-energies of isolated $N_c=4$ clusters.  The hole (electron) doping in DCA calculations is simulated 
in an isolated cluster by just shifting the chemical potential: $\mu<U/2$ ($\mu > U/2$) with constant occupation, $n=1$.  

\begin{figure}
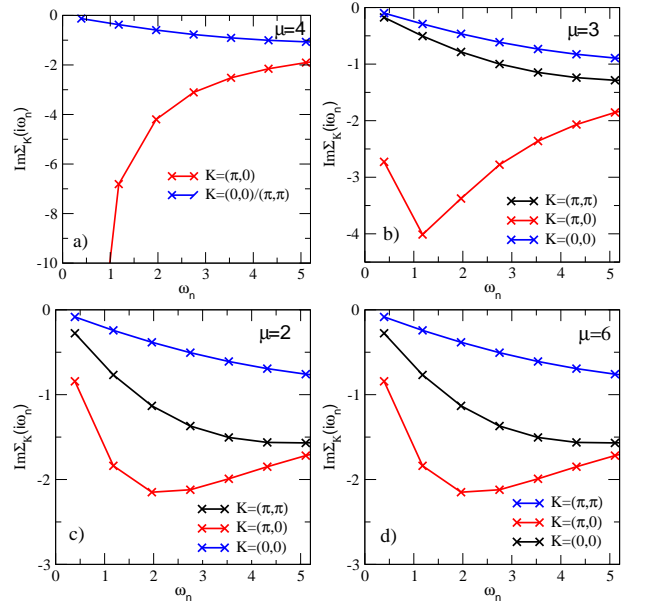

\epsfig{file=fig5a.eps,width=4cm,clip=}
\epsfig{file=fig5b.eps,width=4cm,clip=}
\epsfig{file=fig5c.eps,width=4cm,clip=}
\epsfig{file=fig5d.eps,width=4cm,clip=}
\caption{\label{fig:fig5} (Color online) Dependence of the imaginary part of the self-energy with $\mu$ in isolated $N_c=4$ clusters.  In a) the chemical potential is $\mu=U/2$ 
and there is particle-hole symmetry.  As $\mu$ is shifted downwards for $\mu=3$ in b)  and $\mu=2$ in c), then $|\text{Im}\Sigma_{(\pi,\pi)}(i\omega_n)|$  
is enlarged  with respect to $|\text{Im}\Sigma_{(0,0)}(i\omega_n)|$. In d) we show a case $\mu>U/2$ in which $\mu$ is shifted upwards showing how 
the $(\pi,\pi)$ and the $(0,0)$ self-energies are just switched from the $\mu=2$ situation.  We note that for the $\mu$ values considered,  there is no 
change in the cluster population which is half-filled: $n=1$. The parameters used are $U=8$ and $t=-2$ appropriate for the $N_c=4$ embedded 
DCA calculations in Fig. \ref{fig:fig1}.}
\end{figure}

The imaginary part of the self-energy of an isolated $N_c=4$ cluster is shown in Fig. \ref{fig:fig5}. The $(\pi,0)$ and $(0,\pi)$ sectors display  
divergent behavior: $\text{Im}\Sigma_{(\pi,0)/(0,\pi)}(i\omega_n)\rightarrow -\infty$ as $\omega_n \rightarrow 0$ associated with the 
Mott-Hubbard gap. The imaginary part of the $(0,0)$ and $(\pi,\pi)$ self-energies coincide due to particle-hole symmetry when $\mu=U/2$.
In Fig. \ref{fig:fig5} b) and c)  we show how under a downward shift of $\mu$, they become different:
$|\text{Im}\Sigma_{(\pi,\pi)}(i\omega_n)|> |\text{Im}\Sigma_{(0,0)}(i\omega_n)|$ as found in DCA calculations (see Fig. \ref{fig:fig1}). 
Under an upward shift the behavior in  the $(0,0)$ and $(\pi,\pi)$ self-energies is switched so that: 
$|\text{Im}\Sigma_{(0,0)}(i\omega_n)|> |\text{Im}\Sigma_{(\pi,\pi)}(i\omega_n)|$ as shown in Fig. \ref{fig:fig5} d). 

\begin{figure}
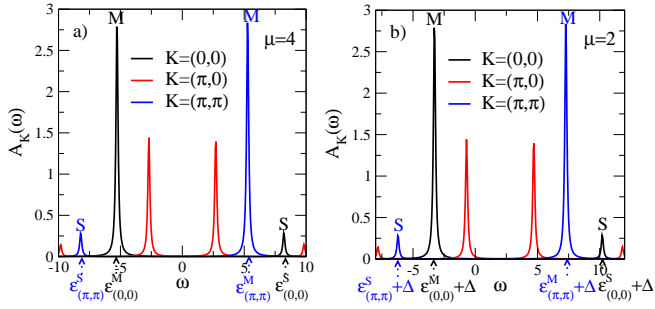

\epsfig{file=fig6a.eps,width=4.1cm,clip=}
\epsfig{file=fig6b.eps,width=4.4cm,clip=}
\caption{\label{fig:fig6} (Color online) Dependence of spectral densities of isolated half-filled $N_c=4$ clusters with the  chemical potential, $\mu$. 
In a) $\mu=U/2$ whereas in b) $\mu<U/2$. The main (M) and satellite (S) peaks of the $(0,0)$ and $(\pi,\pi)$ 
spectral functions are displayed for clarity.  The occupation is at half-filling, $n=1$, in all cases.
The parameters used  are $U=8$ and $t=-2$ and $\Delta=U/2-\mu$. }
\end{figure}

In order to understand the different behavior of the $(0,0)$ and $(\pi,\pi)$ self-energies in the $N_c=4$ isolated cluster 
shown in Fig. \ref{fig:fig5} it is useful to calculate the spectral functions: $A_{\bf K}(\omega)=-{1 \over \pi} \text{Im} G_{\bf K}(\omega+i0^+)$. 
In Fig. \ref{fig:fig6} we show  A$_{\bf K}(\omega)$ for: $\mu=U/2$ which is compared with the: $\mu=2<U/2$ case. 
The main difference between the two cases is a rigid shift of the spectra by $\mu$ with no associated redistribution of weight nor changes in the relative 
peak positions. This is in contrast to the larger enhancement: $|\text{Im}\Sigma_{(\pi,\pi)}(i\omega_n)| > |\text{Im}\Sigma_{(0,0)}(i\omega_n)|$, 
for $\mu=2<U/2$ from which we would have naively interpreted that $(\pi,\pi)$ are more strongly 
correlated than $(0,0)$ electrons. 

The two-peak structure of the spectral density, $A_{\bf K}(\omega)$, 
shown in Fig. \ref{fig:fig6} is adequately described through a single-pole description 
of the self-energy:
\begin{equation}
\Sigma_{\bf K}(i \omega_n)-U { n \over 2} =  {E_{\bf K}  \over i \omega_n -\Delta- F_{\bf K}},
\label{eq:selfpole}
\end{equation}
with $\Delta={U n \over 2}-\mu$ and the self-energy pole position $F_{\bf K}$ and the constant $E_{\bf K}$ independent of $\mu$. 
We use $n=1$ in the Hartree contribution to the self-energy since we are at half-filling. 

By introducing Eq. (\ref{eq:selfpole}) in the Greens function of Eq. (\ref{eq:green}) and performing analytical continuation to the
real axis:
\begin{eqnarray}
A_{\bf K}(\omega)=-{1 \over \pi} ImG_{\bf K}(\omega+i0^+)
\nonumber \\
= \left|{\epsilon^M_{\bf K}-F_{\bf K} \over \epsilon^M_{\bf K}-\epsilon^S_{\bf K}} \right| \delta(\omega-\Delta-\epsilon^M_{\bf K})
+\left|{\epsilon^S_{\bf K}-F_{\bf K} \over \epsilon^S_{\bf K}-\epsilon^M_{\bf K} } \right| \delta(\omega-\Delta-\epsilon^S_{\bf K}),
\label{eq:Ak}
\end{eqnarray} 
with the location of the two peaks given by:
\begin{eqnarray}
\epsilon^M_{\bf K}&=&{ \epsilon_{\bf K}+F_{\bf K} \over 2} - \sqrt{ \left({\epsilon_{\bf K}-F_{\bf K} \over 2} \right)^2 + E_{\bf K}},
\nonumber \\
\epsilon^S_{\bf K}&=&{ \epsilon_{\bf K}+F_{\bf K} \over 2} + \sqrt{ \left({\epsilon_{\bf K}-F_{\bf K} \over 2} \right)^2 + E_{\bf K}},
\label{eq:peaks}
\end{eqnarray}
for ${\bf K}=(0,0)$ whereas the $M$ and $S$ peak labels are switched for ${\bf K}=(\pi,\pi)$.  The analytical expression (\ref{eq:Ak}) for $A_{\bf K}(\omega)$ 
shows a two-peak structure as expected. Since the weights of the delta peaks in Eq. (\ref{eq:Ak}) are independent of $\Delta$,
it is evident that a shift in $\mu$ just leads to a rigid shift with no redistribution of the spectrum as found in the isolated cluster. 
The locations and weights of the two peaks occurring in $A_{(0,0)}(\omega)$ and $A_{(\pi,\pi)}(\omega)$ 
shown in Fig. \ref{fig:fig6} are faithfully reproduced by using: $F_{(0,0)}=-F_{(\pi,\pi)}=6.95$, and $E_{(0,0)}=E_{(\pi,\pi)}=15.25$ in the
self-energy of Eq. (\ref{eq:selfpole}). These parameters are obtained using the
exact peak locations for $\epsilon^M_{\bf K}$ and $\epsilon^S_{\bf K}$ in Eq. (\ref{eq:peaks})

We finally analyze how the shift in $\mu$ modifies the $(0,0)$ and $(\pi,\pi)$ self-energies based on the single-pole form.
The imaginary part of the self-energy in Eq. (\ref{eq:selfpole}) reads:
\begin{equation}
\text{Im}\Sigma_{\bf K}(i\omega_n)=-{E_{\bf K}  \over (\Delta+F_{\bf K})^2 + \omega_n^2 }\omega_n.
\label{eq:imsim}
\end{equation} 
In the symmetric case, $\mu=U/2$ ($\Delta=0$), we have that: Im$\Sigma_{(0,0)}(i\omega_n)$=Im$\Sigma_{(\pi,\pi)}(i\omega_n)$ since the self-energy 
poles are symmetrically located: $|\Delta + F_{(0,0)}|=|\Delta+F_{(\pi,\pi)}|$.  However, if $\mu$ is shifted downwards ($\Delta=U/2-\mu>0$) 
then $|\Delta+F_{(\pi,\pi)}|$ ($|\Delta+F_{(0,0)}|$) is suppressed (enhanced). From Eq. (\ref{eq:imsim}),  this leads to an 
enhancement of $|\text{Im}\Sigma_{(\pi,\pi)}(i\pi/\beta)|$ and a suppression of $|\text{Im}\Sigma_{(0,0)}(i\pi/\beta)|$ 
with respect to their $\Delta=0$ values. Taking the $F_{\bf K}, E_{\bf K}$ parameters used 
above to reproduce the spectral densities of Fig. \ref{fig:fig6}, the self-energy behavior of Fig. \ref{fig:fig5} by which
 $|\text{Im}\Sigma_{(\pi,\pi)}(i\pi/\beta)|> |\text{Im}\Sigma_{(0,0)}(i\pi/\beta)|$ under a downward shift of $\mu$ is correctly 
reproduced. We note that the single pole functional form of Eq. (\ref{eq:selfpole}) has been previously introduced  
in single site DMFT \cite{kotliar1996} for analyzing the atomic limit of the Mott insulator and more recently
for interpreting particle-hole asymmetries in the electronic properties of the doped Hubbard 
model.\cite{kotliar2013}.  
 \begin{figure}
\epsfig{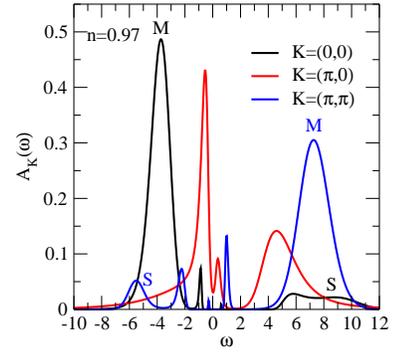}
\caption{\label{fig:fig7} (Color online) Spectral densities obtained from DCA on an $N_c=4$ cluster at low hole dopings. 
The parameters used here are $U=8,\beta=8, t=-1$ with $\mu=2.2$. }
\end{figure}

It is now worth analyzing the relevance of the single-pole self-energy functional of Eq. (\ref{eq:selfpole}) 
to embedded clusters. Does the electronic structure of $(0,0)$ and $(\pi,\pi)$ approximately  
behaves as in isolated clusters discussed above? To answer this question, we have obtained the DCA spectral functions 
at low hole dopings which are shown in Fig. \ref{fig:fig7}. The $(\pi,0)$ spectral function displays a 
pseudogap at the chemical potential in agreement with previous works \cite{jarrell2005,civelli2005,gull2010,merino2014,gunnarsson2015} which 
we don't discuss further here. The $(0,0)$ and $(\pi,\pi)$ spectral functions are essentially gapped and mainly consist of a main peak 
containing most of the spectral weight and a satellite peak with much more smaller weight. 
The main effect of the downward  chemical potential shift, $\mu=2.2<U/2$, which slightly hole dopes the system, $n=0.97$, is to  
rigidly shift $A_{\bf K}(\omega)$ such that the main peak in $A_{(0,0)}(\omega)$ becomes much closer to the chemical potential 
than the main peak in $A_{(\pi,\pi)}(\omega)$.  Hence, the overall behavior of $A_{(0,0)}(\omega)$ and $A_{(\pi,\pi)}(\omega)$ 
in DCA  at low hole dopings is consistent with a rigid upward shift  of the main spectral function structures similarly to the 
overall behavior found in isolated clusters.
There are, however, some features in $A_{(0,0)}(\omega)$ and $A_{(\pi,\pi)}(\omega)$ intrinsic to DCA spectral functions not found in isolated clusters. 
The main peak in $A_{(\pi,\pi)}(\omega)$ is more broad as compared to the main peak
in $A_{(0,0)}(\omega)$ when $\mu<U/2$. This behavior is reasonable since 
there are more decay possibilities for electrons excited further away from the Fermi surface. Apart from this broadening 
there are some smaller structures occurring around the chemical potential in the DCA which do not occur in isolated clusters.  
Due to the small weight of these features, the overall behavior of the $(0,0)$ and $(\pi,\pi)$ self-energies is 
dominated by the position of the main peaks in $A_{(0,0)}(\omega)$ and $A_{(\pi,\pi)}(\omega)$ in consistent
agreement with the isolated cluster analysis.

\section{Conclusions and discussion}
\label{sec:conclusion}
In the present work we have analyzed the doping dependence of DCA self-energies, $\Sigma_{\bf K}(i\omega_n)$, 
for ${\bf K}$ away from the Fermi surface in a Hubbard model 
on the square lattice.  For hole doping,  we find larger enhancements of the self-energy magnitudes
at the top (${\bf K}=(\pi,\pi)$) than at the bottom (${\bf K}=(0,0)$) 
of the band which would naively imply that electron correlation effects are stronger at 
$(\pi,\pi)$ than at $(0,0)$. However,  the DCA populations satisfy: $n_{(0,0)\sigma} \approx 1-n_{(\pi,\pi)\sigma}$, which would
naively suggest similar electron correlation effects for $(0,0)$ and $(\pi,\pi)$ momenta. 

In order to understand the origin of such self-energy differences we have first clarified 
the role played by bath-cluster hybridizations. Our DCA analysis shows 
that  the self-energy difference: $|\Sigma_{(\pi,\pi)}(i\omega_n)|>|\Sigma_{(0,0)}(i\omega_n)|$ is not related to
the $(0,0)$ hybridization being stronger than the $(\pi,\pi)$ hybridization.  Indeed, such self-energy difference
is also found in half-filled ($n=1$) isolated clusters but with $\mu<U/2$.  Based on the equivalent spectral weight distributions 
at ${\bf K}=(0,0)$ and ${\bf K}=(\pi,\pi)$, we conclude that electron correlation effects acting at ${\bf K}=(\pi,\pi)$ and at $(0,0)$ should be
similar in the isolated cluster. This is in contrast to the conclusion we would have reached by just 
looking at the larger self-energy enhancement at $(\pi,\pi)$ compared to $(0,0)$. 

Our analysis indicates that DCA self-energies at ${\bf k}$-points far away from the Fermi surface 
contain "apparent" enhancements which do not  necessarily correspond to electronic correlation effects. 
These enhancements are a consequence of shifting the chemical potential from $\mu=U/2$ in a Mott insulator 
with the nearly constant occupation, $n \approx 1$.
From an experimental point of view it would be interesting to compare 
the spectral functions of electrons close to the bottom of the band with electrons at the top of the band
in hole doped cuprates. This would require probing unoccupied electronic states using 
angular resolved inverse photoemission\cite{ARIPES} (ARIPES) in combination with the more popular
ARPES \cite{ARPES} experiments probing occupied states.
\section*{Acknowledgements.} 
J.M.  acknowledges financial support from MINECO (MAT2012-37263-C02-01).

\end{document}